\documentclass{article}
\usepackage{amsmath,amssymb,amscd,latexsym,amsthm}
\usepackage{graphicx}
\usepackage{slashed}
\usepackage{hyperref}
\usepackage[backend=bibtex]{biblatex}
\usepackage{tikz}
\usepackage{tikz-cd}
\author{Peter Woit \\
Department of Mathematics\\
Columbia University}
\title{Is Space-time Really Doomed?} 
\begin{document}
\maketitle
\begin{abstract}
For many years now it has become conventional (\cite{witten},\cite{gross},\cite{arkani-hamed}) for theorists to argue that \lq\lq space-time is doomed", with the difficulties in finding a quantum theory of gravity implying the necessity of basing a fundamental theory on something quite different than usual notions of space-time geometry.  But what is this space-time geometry that is doomed?  In this essay we'll explore how our understanding of four-dimensional geometry has evolved since Einstein, leading to new ideas about such geometry which may not be doomed at all.

\end{abstract}
%\doublespace%
%Essay written for the Gravity Research Foundation 2022 Awards for Essays on Gravitation.
%\newpage

\section{Geometry in terms of metrics}

The traditional formulation of geometry used by Einstein in his discovery of general relativity goes back to Riemann.   Its modern incarnation is well-represented in 
the textbook \cite{mtw} and is based upon the following elements:
\begin{itemize}
\item A four-dimensional manifold $M$ described in terms of coordinate charts given by a covering of $M$ by $U_j\subset M$ and maps
$$\phi_j :U_j \rightarrow \mathbf R^4 $$
This manifold is usually taken to be smooth ($\phi_k(\phi_j^{-1})$ is $C^\infty$ where defined).
\item Tensor fields, which at $m\in M$ take values in a tensor product of copies of the tangent ($T_mM$) or cotangent ($T^*_mM$) space. Using the coordinate charts these are given by functions with some number of upper and lower indices.
\item The metric tensor field $g$ takes values in the symmetric subspace of $T^*_mM\otimes T^*_mM$ and has signature $(3,1)$.
\end{itemize}
The metric tensor determines a connection, the Levi-Civita connection, and Einstein's equations are written in terms of the curvature of this connection (the Riemannian curvature) and the energy-momentum tensor.   One can derive Einstein's equations as Euler-Lagrange equations of the Einstein-Hilbert action given by the integral of the scalar curvature.

The data used to describe the geometry is highly redundant, due to the freedom to choose coordinate charts.  In a Hamiltonian formalism where initial value data is given by a metric on a $3d$ hypersurface and its time derivative, configuration space is $6$ dimensional (the metric has $6$ coordinates).  Diffeomorphism invariance implies freedom to make coordinate changes in each of four dimensions and thus four constraints, so $2$ physical degrees of freedom.

\section{Geometry in terms of frame bundles with connection}

Soon after Einstein's discovery of general relativity, Cartan gave a different formulation of geometry, which after later rewriting in the language of principal bundles is well-described in the textbook \cite{kobayashi-nomizu}.   Here the geometry is described by
\begin{itemize}
\item A $10$-dimensional principal $SO(3,1)$-bundle over a $4$-dimensional space-time, the bundle $OF(M)$ of orthonormal frames.
\item A $1$-form  $\omega$ on $OF(M)$ taking values in the Lie algebra of $SO(3,1)$, the spin connection.
\item An $\mathbf R^4$-valued $1$-form $e$ on $OF(M)$, the tetrad.
\end{itemize}
Writing the Einstein-Hilbert action in terms of $e, \omega$ and its curvature $\Omega$ the action is
\begin{equation}
\label{eq:action}
\int_M \epsilon_{ABCD}e^A\wedge e^B\wedge \Omega^{CD}(\omega)
\end{equation}
Varying $\omega$ gives an equation of motion setting the torsion to zero (with a unique solution for $\omega$ in terms of $e$, the Levi-Civita connection).  Varying $e$ gives the Einstein equations.  

Using these variables, in the Hamiltonian formalism the configuration space of orthonormal frames on a 3d hypersurface is $9$-dimensional.  In addition to the four constraints coming from diffeomorphism invariance, there are now three more coming from freedom to rotate the frames by an element of $SO(3)$, again leaving $2$ physical degrees of freedom.

One ends up with the same field equations, but two important advantages of Cartan's formulation are:
\begin{itemize}
\item One can describe not just tensor fields, but also spinor fields, by taking the spin double-cover of $OF(M)$, which has fiber $SL(2,\mathbf C)$ rather than $SO(3,1)$.
\item Gauge theory can be formulated in the same language, by taking an arbitrary Lie group $G$ and principal $G$-bundle over $M$, with an arbitrary connection $A$, which will be a $1$-form valued in the Lie algebra of $G$.  One has a Yang-Mills rather than Einstein-Hilbert action, given by the norm-squared of the curvature of $A$.
\end{itemize}
The first of these is important since matter fields in nature are spinor fields, the second because one ultimately would like to have a unified framework for describing general relativity and the Standard Model.

\section{Frame bundles with connection in four dimensions}

The metric formalism and the Cartan tetrad/connection formalism can be used to describe geometry in the same way in any dimension.   In four dimensions the tetrad/connection formalism has special features, due to the fact that the Hodge $*$-operator takes $2$-forms to $2$-forms.   This operator satisfies $*^2=1$ for Euclidean signature, so one can break up two-forms into self-dual ($*=1$) and anti-self-dual ($*=-1$) subspaces.   This implies that the Lie algebra of the rotation group is not simple, but decomposes as
$$\frak{so}(4)=\frak{so}(3)\oplus \frak{so}(3)=\frak{su}(2)\oplus \frak{su}(2)$$
In Minkowski signature one has $*^2=-1$, so such a decomposition requires complexifying the Lie algebra to get eigenspaces with eigenvalues $\pm i$, giving
$$\frak{so}(3,1)\otimes \mathbf C=\frak{sl}(2,\mathbf C)\oplus \frak{sl}(2,\mathbf C)$$

Remarkably, one can describe general relativity in four dimensions using only half the connection variables needed in other dimensions (see for instance \cite{krasnov}).  Using just the self-dual component  to compute the curvature in the action (in the form of equation \ref{eq:action}) one gets the same Einstein equations and same counting of number of degrees of freedom.    This is straight-forward in Euclidean signature, but in Minkowski signature one needs to complexify (doubling the number of degrees of freedom) in order to project out the self-dual component, and then later impose a reality condition as a constraint.

Ashtekar variables \cite{ashtekar} provide a phase-space realization of this decomposition, and a starting point for alternate quantization methods such as loop quantum gravity. In Minkowski signature,  the need to work with complexified connections and impose reality conditions leads to significant problems which need to be overcome.

\section{Twistor geometry}

Again special to four-dimensions is a more radical change in the description of space-time due to Roger Penrose \cite{penrose}, who proposed to take points in space-time to be two-dimensional complex subspaces of a space $\mathbf C^4$ which is called twistor space.  Taking all such subspaces, one gets not Minkowksi space-time, but a complexified and conformally compactified Minkowksi space-time.   The usual tensor field description of geometry has no way to describe spinor fields, while in the Cartan formalism there is a way to describe them, but they are not especially simple or well-motivated.  By contrast, in the twistor framework, chiral spinors are tautological objects:  the two-dimensional chiral spinor space at a point is nothing but the point itself.

Another significant advantage of the twistor framework is that conformal symmetry has a simple description.  The group $SL(4,\mathbf C)$ acting linearly on twistor space is the complexification of the conformal group.

A characteristic feature of twistor geometry is that it most simply describes not Euclidean or Minkowski space-time, but something which is simultaneously a complexification of both of these.  This allows an explicit understanding of how to pass between Euclidean and Minkowski by analytic continuation.  The conformal groups $Spin(5,1)$ (Euclidean) and $Spin(4,2)$ (Minkowski) are different real forms of the complex conformal group $SL(4,\mathbf C)=Spin(6,\mathbf C)$.  These have as subgroups the real forms $Spin(4)=SU(2)\times SU(2)$ and $Spin(3,1)=SL(2,\mathbf C)$ of $Spin(4,\mathbf C)$.

\section{Euclidean quantum field theory}

Attempts to formulate quantum field theories on Minkowski space-time suffer from a variety of serious mathematical difficulties.  Even in the case of the simplest free quantum field theories, the two-point function is not a function but best defined as a hyperfunction: one needs to go to complexified space-time and then take boundary values of holomorphic functions.  In Euclidean space-time on the other hand, the same free field two-point function is a well-behaved function and can be rigorously defined using path integrals (for more about this, see for instance \cite{jaffe}).   The only known non-perturbative definition of Yang-Mills theory is in Euclidean space-time.  It is reasonable to conjecture that all fundamental physical theories need to be based on a Euclidean space-time definition, with Minkowski space-time amplitudes only recovered by taking boundary values of analytic continuations from Euclidean space-time.

While amplitudes can be analytically continued, Euclidean fields can't.  Such fields are quite different objects than Minkowski space fields: they are always \lq\lq off-shell", not satisfying any equations of motion.  While there is a Euclidean Fock space formalism, this is not at all the same as the physical Fock space formalism that describes multi-particle systems.   Another way in which Euclidean quantum field theory differs from Minkowski quantum field theory is that one needs to break $SO(4)$ rotational invariance and pick a specific imaginary time direction.  Only once this is done can one define physical states and the analytic continuation to Minkowski space-time.

\section{Euclidean twistors and unification}

The recent preprint \cite{woit} describes a speculative framework that uses twistors and Euclidean quantum field theory to put together the elements of a gravity theory in a chiral formulation as described above with the degrees of freedom of the Standard Model.   While the connection for one of the $SU(2)$ factors in $Spin(4)$ provides the chiral spin connection for gravity, the other $SU(2)$ factor behaves like an internal symmetry, providing the gauge fields of the weak interactions.   The Higgs field which spontaneously breaks this second $SU(2)$ is the field needed to break $SO(4)$ rotational invariance and pick out a specific imaginary time direction.  While much remains to be done to turn this framework into a full theory with well-defined dynamics, the degrees of freedom and symmetries of a unified theory are in place, in a way that does not seem to have been previously studied.

\section{Conclusions}

The fundamental variables of the Standard Model are geometrical, with forces described in terms of connections and curvature.  It is often assumed that the Standard Model is just an effective low energy theory, but with the Yang-Mills dynamics the quantum theory appears to be consistent at arbitrarily short distances (for the $U(1)$ gauge theory a potential problem only appears at scales far smaller than the Planck scale).   There is no reason to believe that the connection/curvature/spinor geometry of the Standard Model is \lq\lq doomed" and needs to be replaced at short distances.   Equally importantly, there is no known viable theoretical framework  that would replace connections/curvatures/spinors at short distances in terms of something very different.

The close relation of the mathematical structures of four-dimensional geometry to the connections/curvatures/spinors of the Standard Model indicate that any attempt to abandon these structures in favor of something completely different will face insurmountable problems.  How will some completely different framework manage to unify with and reproduce the successes of the Standard Model?  We argue that four-dimensional geometry (in a formulation that brings Euclidean signature, spinors and twistors to the fore) provides exactly the needed degrees of freedom and symmetries for a unified theory of space-time gravitation and known particle physics.  This gets the kinematics  correct, with the problem remaining that of dynamics.   Yang-Mills theory shows that consistent short-distance dynamics of gauge fields does exist, and twistor variables provide a natural way to get short-distance conformal invariance.   Perhaps one can find consistent short-distance dynamics by working with a chiral formulation for gravity in the twistor framework, taking as fundamental Euclidean signature with its special role for the imaginary time direction.   Likely what is doomed at short distances is not space-time geometry, but only the Einstein-Hilbert action, which may be just a long-distance effective action.

%\singlespace
\printbibliography[title={References}]
\end{document}